\documentclass[11pt]{article}
\usepackage{amsfonts}
\usepackage{epic}
\newcommand{\be}{\begin{equation}\label}
\newcommand{\ee}{\end{equation}}
\newcommand{\bea}{\begin{eqnarray}\label}
\newcommand{\eea}{\end{eqnarray}}

\newcommand{\tr}{\tilde{\rho}}

\begin{document}
\begin{center}

{\Large \textsf{Thermodynamic Bethe Ansatz  for the\\[4mm]
Spin-1/2 Staggered XXZ- Model }}

\vspace{36pt}

{\bf V.V.Mkhitaryan}\footnote{e-mail:{\sl vgho@lx2.yerphi.am}} and
{\bf A.G.Sedrakyan}\footnote{e-mail:{\sl sedrak@lx2.yerphi.am}}

\vspace{5pt}

{\small Yerevan Physics Institute, Armenia}

\vfill

{\bf Abstract}

\end{center}

We develop the technique of Thermodynamic Bethe Ansatz to
investigate the ground state and the spectrum in the thermodynamic
limit of the staggered $XXZ$ models proposed recently as an
example of integrable ladder model. This model appeared due to
staggered inhomogeneity of the anisotropy parameter $\Delta$ and
the staggered shift of the spectral parameter. We give the structure of ground states and
lowest lying excitations in two different phases which occur at zero temperature.

 \vfill


\rightline{January 2003}


\newpage

\section{Introduction}

\noindent

The quasi-one-dimensional spin ladder models have attracted much attention in recent years; 
besides the mathematical interest, these models are believed to describe many real magnetic materials invented
relatively recently \cite{hiroiazuma}.

In connection with this  the construction and exact solution of
exactly integrable ladder models became very actual. In particular the Bethe
Ansatz exact solution can provide a deep insight into the properties of these models.
Recently in a series of articles \cite{ararub, AASSS,ST, ASSS}, it have been proposed a technique for constructing 
integrable models where the model parameter $\Delta$ (the anisotropy
parameter of the $XXZ$ and anisotropic $t-J$ models) have a
staggered disposition of the sign along both chain and time
directions. Due to the staggered shift of the spectral parameter
also considered there, these models have led to Hamiltonians
formulated on two leg zig-zag quasi-one dimensional chains.

The technique is based on an appropriate modification of the
Yang-Baxter equations ($YBE$) (\cite{Bax, FT, korepin}) for the
$R$-matrices which are the conditions for commutativity of the
transfer matrices at different values of the spectral parameter.
The transfer matrix is defined as the staggered product of
$R$-matrices, which have staggered sign of the anisotropy
parameter $\Delta$ and staggered shift of the spectral parameter
of the model along the chain. The shift of spectral parameter in a
product of $R$-matrices was considered in earlier works of several
authors \cite{SW, dV}, by use of which they have analyzed the
ordinary $XXZ$ model in the infinite limit of the spectral
parameter. Later, in the article \cite{zvyagin}, this modified
transfer matrices has been used in order to construct integrable
models on the ladder.

Various integrable models on ladders were also constructed in 
articles \cite{Wa, Ko, Muu, Al, For}. The above described construction
essentially differs from those due to the
inhomogeneity of anisotropy parameter along the chain.

The subject of present paper is to study the exactly soluble spin-1/2 two-leg zig-zag ladder model proposed 
in \cite{ararub}, in the thermodynamic limit. We use the technique of Thermodynamic
Bethe Ansatz developed by M. Gaudin \cite{gaudin} and M.Takahashi and M.Suzuki \cite{takahashi}  for the Heisenberg 
$XXZ$ model. This method is based on the study of Bethe Ansatz equations (BAE). Another powerful method
in theoretical study of spin ladder models which we partially use is the bosonization method \cite{gnt,shn,ne}.   
The bosonization have been used very often to find out and to understand many physical aspects of these systems . 
This method yields a low-energy effective field theory of a spin system, which can be treated by powerful methods of
quantum field theory.

\begin{figure}[h]
  \begin{center}
    \leavevmode
\setlength{\unitlength}{0.0015cm}
\begingroup\makeatletter\ifx\SetFigFont\undefined
\def\x#1#2#3#4#5#6#7\relax{\def\x{#1#2#3#4#5#6}}%
\expandafter\x\fmtname xxxxxx\relax \def\y{splain}%
\ifx\x\y   
\gdef\SetFigFont#1#2#3{%
  \ifnum #1<17\tiny\else \ifnum #1<20\small\else
  \ifnum #1<24\normalsize\else \ifnum #1<29\large\else
  \ifnum #1<34\Large\else \ifnum #1<41\LARGE\else
     \huge\fi\fi\fi\fi\fi\fi
  \csname #3\endcsname}%
\else
\gdef\SetFigFont#1#2#3{\begingroup
  \count@#1\relax \ifnum 25<\count@\count@25\fi
  \def\x{\endgroup\@setsize\SetFigFont{#2pt}}%
  \expandafter\x
    \csname \romannumeral\the\count@ pt\expandafter\endcsname
    \csname @\romannumeral\the\count@ pt\endcsname
  \csname #3\endcsname}%
\fi
\fi\endgroup
{\renewcommand{\dashlinestretch}{30}
\begin{picture}(7954,2733)(0,-10)
\drawline(462,2304)(1362,504)(2262,2304)
        (3162,504)(4062,2304)(4962,504)
        (5862,2304)(6762,504)
\drawline(12,2304)(7212,2304)
\drawline(12,504)(7212,504)
\put(1182,54){\makebox(0,0)[lb]{\smash{{{\SetFigFont{12}{14.4}
{\rmdefault}$j$}}}}}
\put(2982,54){\makebox(0,0)[lb]{\smash{{{\SetFigFont{12}{14.4}
{\rmdefault}$j+1$}}}}}
\put(4782,54){\makebox(0,0)[lb]{\smash{{{\SetFigFont{12}{14.4}
{\rmdefault}$j+2$}}}}}
\put(2082,2574){\makebox(0,0)[lb]{\smash{{{\SetFigFont{12}{14.4}
{\rmdefault}$j$}}}}}
\put(3882,2574){\makebox(0,0)[lb]{\smash{{{\SetFigFont{12}{14.4}
{\rmdefault}$j+1$}}}}}
\put(7662,2304){\makebox(0,0)[lb]{\smash{{{\SetFigFont{12}{14.4}
{\rmdefault}$s=1$}}}}}
\put(7662,504){\makebox(0,0)[lb]{\smash{{{\SetFigFont{12}{14.4}
{\rmdefault}$s=0$}}}}}
\end{picture}
}
\end{center}
\caption{Zig-zag ladder chain}
\label{fig1}
\end{figure}
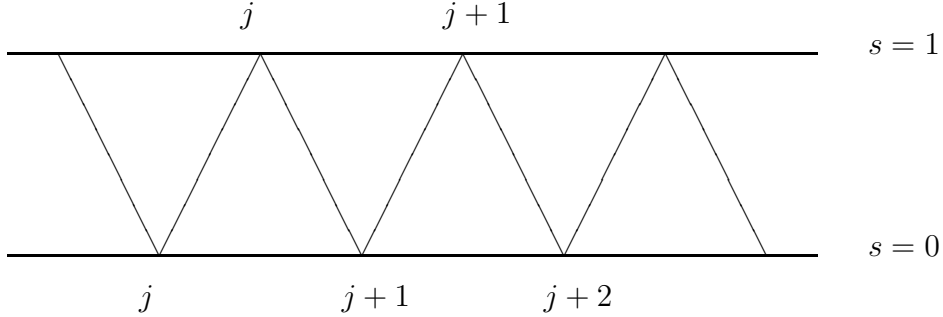

In the model under consideration, the $R$-matrix is a product of two samples of XXZ $R$- matrices with the staggered sign of anisotropy
parameter $\Delta$ and the shift of spectral parameter by a new model parameter. The Hamiltonian 
\bea{spinham}
H&&\hspace{-1cm}=\sum\limits_{j=1}^N\sum\limits_{s=0,1}\left \{{(-1)^s}/2 \left[  \sigma_{j,s}^x \sigma_{j+1,s}^x+  \sigma_{j,s}^y \sigma_{j+1,s}^y-  \sigma_{j,s}^z \sigma_{j+1,s}^z\right]\right .\nonumber\\
&&\,\nonumber\\
+ &J_1& (-1)^s\left[ \sigma_{j,s}^z( \sigma_{j,s+1}^+ \sigma_{j+1,s}^- -  \sigma_{j,s+1}^- \sigma_{j+1,s}^+) -
\sigma_{j+1,s}^z( \sigma_{j,s}^+ \sigma_{j,s+1}^- -  \sigma_{j,s}^- \sigma_{j,s+1}^+) \right]\nonumber\\
&&\,\nonumber\\
-&J_2&\left .\left[ \sigma_{j,s+1}^z( \sigma_{j,s}^+ \sigma_{j+1,s}^- -  \sigma_{j,s}^- \sigma_{j+1,s}^+)\right]
+\mu_0H\sigma^z_{j,s}\right \};\nonumber
\eea
$$
\sigma_{N+1,s}\equiv  \sigma_{1,s},\qquad  \sigma_{j,2}\equiv  \sigma_{j+1,0},\qquad J_1=\frac{\sin\lambda}{\sin\theta},\qquad J_2=\tan\lambda\cot\theta.
$$
describes the interaction of $2N$ spins located on the sites of zig-zag ladder (see Fig.\ref{fig1}), in the
external "magnetic" field $H$ along the $z$-axis.
Parameterizations of coupling constants $J_1$ and $J_2$ are given by the two combinations
\bea{par1}
&&1.\,\lambda\sim\mbox{real},\quad\theta\sim\mbox{imaginary},\quad\mbox{then}\,\,0\leq |J_1|\leq |J_2|\leq\infty,\\
&&\,\nonumber\\
&&2.\,\lambda\sim\mbox{imaginary},\quad\theta\sim\mbox{real},\quad\mbox{then}\,\,0\leq
|J_2|< |J_1|\leq\infty.
\label{par2}
\eea
which fix the Hamiltonian to be Hermitian and give all the possible (imaginary) values of $J_1$ and
$J_2$.

By construction, this model can be solved exactly through the Bethe Ansatz technique \cite{ararub}.
Starting from the ferromagnetic eigenstate with all spins up, each $M$- particle eigenstate is obtained
by adding $M$ magnons which are parameterized by unequal complex rapidities $x_1,..,x_M$, the roots of
the Bethe Ansatz equations \cite{ararub}
\be{BAE}
\left[\frac{\sinh\left(\frac\lambda 2(x_k-i\frac{\theta}\lambda-i)\right)\sinh\left(\frac\lambda 2(x_k+
i\frac{\theta}\lambda-i)\right)}
{\sinh\left(\frac\lambda 2(x_k-i\frac{\theta}\lambda+i)\right)\sinh\left(\frac\lambda 2(x_k+i\frac{\theta}\lambda+i)\right)}\right]^N=-\prod\limits_{i=1}^M
\frac{\sinh\left(\frac\lambda 2(x_k-x_i-2i)\right)}{\sinh\left(\frac\lambda 2(x_k-x_i+2i)\right)}.
\ee
The corresponding state has total spin projection $S_z=N-M$ and energy
\bea{selfenergy}
&&\hspace{-1cm} E(x_1,..,x_M)=\frac{\sin(\lambda+\theta)\sin(\lambda-\theta)\sin\lambda}{\sin^2\theta\cos\lambda}\times\\
&&\hspace{-0.5cm}\sum\limits_{j=1}^M\left\{\frac {\sin\lambda}{\sinh\left(\frac\lambda 2(x_j+i\frac{\theta}\lambda-i)\right)
\sinh\left(\frac\lambda 2(x_j+i\frac{\theta}\lambda+i)\right)} -
\frac {\sin\lambda}{\sinh\left(\frac\lambda 2(x_j-i\frac{\theta}\lambda-i)\right)
\sinh\left(\frac\lambda 2(x_j-i\frac{\theta}\lambda+i)\right)}\right\}.\nonumber
\eea
This type of BAE arise also in the study of lattice limits of one-dimensional field theoretical
models \cite{japaners, dut, devega, destri}, as well as in some exactly solvable zig-zag ladder models
considered in \cite{zvyagin, frahm}. But the similarity is restricted only by the Bethe equations,
because for all of these models the bare energy $h(x)$ is even function of the rapidity $x$ in contrast of
our case, where $h$ is apparently odd function of $x$. As we shall see, this makes the physical
pictures very different.

The paper is organized as follows. In section 2 the BAE and the classification of their solutions in different sectors 
of the model are given, in the thermodynamic limit $N\rightarrow\infty$.
The spectral equations provided that the system is in the statistical equilibrium state are derived. In section 3, the
spectral equations are considered in zero "magnetic" field $H=0$. We have reduced the system of spectral equations in 
zero temperature limit $T\rightarrow 0$ to a single integral equation
and sketched out the structure of ground states and lowest lying excitations
of the model, for two different sectors (\ref{par1}) and (\ref{par2}). We shown that the
spectrum is gapless in (\ref{par1}) and has both gapped and gapless excitations in (\ref{par2}).
Two special limits $J_1=0,\,J_2\neq 0$ and $J_1=J_2$
are considered in section 4. We have shown that when $J_1=0$, in the thermodynamic limit
our model is equivalent to two species of Heisenberg $XXZ$ models with different signs of 
hopping constants. We studied the interaction between the two species, which is of next to 
leading order, by use of both the finite size correction and the bosonization methods. We
recognized that the limit $J_1=J_2$ inherits some features from both $|J_1|<|J_2|$ and
$|J_1|>|J_2|$ regions. We have summarized our results in the last section.

\section{The Bethe Ansatz equations in the thermodynamic limit and the spectral equations}

In what follows we shall be interested in the thermodynamic limit of the theory. Namely, the size of
the system $N$ and the number of elementary excitations in the energy interval $[\varepsilon,\,\varepsilon+d\varepsilon]$,
$M(\varepsilon)$, will go to infinity by the finite ratio

$$ N\rightarrow\infty,\quad M(\varepsilon)\rightarrow\infty,\quad M(\varepsilon)/N=\mbox{const}
$$
Then, according to \cite{gaudin, takahashi}, general solutions of (\ref{BAE})
are grouped in so-called strings. The character of the string solution strongly depends on whether the proper value of complex
parameter $\lambda$ is real or imaginary. Thus in two regions (\ref{par1}) and (\ref{par2}) the
model exhibits different properties which we shall analyze separately.

In spite of possible shortcomings of the string hypothesis ( see for instance \cite{korepinessler} ), we base our 
analysis on the string solution approach, which we believe renders correct results in this thermodynamic regime, as 
it is in the case of Heisenberg spin chains. 

It is worth to mention that we take the parameter $\theta$ to be finite. 
In the absence of magnetic field, all the spectral functions typically go to zero at 
(finite or infinite) Fermi points. Then we expect that in the region (\ref{par1}) there 
will be no effective restrictions on the variation range of spectral parameters, and the 
spectrum will be gapless.
 
In this section we shall bring the basic properties of Bethe Ansatz exact solution. This
solution will allow to derive the spectral equations of the theory which describe the spectrum
of physical excitations and provide a systematic method for constructing the physical vacuum
and the equilibrium properties of the system.

\subsection{The case of real $\lambda$ and imaginary $\theta$ $(|J_1|\leq|J_2|)$}
\subsubsection{The properties of Bethe Ansatz equations}
In this case we change the parameter $\theta\rightarrow i\theta$ in order to deal with real $\theta$.

Equations (\ref{BAE}) and eigenvalues (\ref{selfenergy}) have periodic property with respect to the shift of parameters
$x_i\rightarrow x_i+2p_0 i$, where the parameter $p_0=\frac \pi\lambda$ is introduced. In the thermodynamic limit, string solutions have the form \cite{takahashi}
\be{strings1}
x^{n,k}_{\alpha,\pm}= x^{n}_{\alpha,\pm}+i(n+1-2k)+i\frac{p_0}2(1-v),\quad (\mbox{mod}\,\,2p_0i),\quad
k=1,2,..,n;
\ee
characterized by the parity $v=\pm 1$, the real part ( the center of string ) $x^{n}_{\alpha,\pm}$ and the order $n$. The allowed values of $n$
are restricted by the conditions
\be{restrn}
v\sin[k\lambda]\sin[(n-k)\lambda]>0,
\ee
which follows from the normalizability conditions for the wave functions \cite{zotos}.

Due to the symmetry $\lambda\rightarrow\pi-\lambda$ of the eqs. (\ref{BAE}), we shall first consider
the case $\lambda<\pi/2$. Solutions of (\ref{restrn}) for arbitrary values of parameter $p_0$
are given in \cite{takahashi}. The starting point is the representation of $p_0$ in terms of
continued fraction
\be{confrac}
p_0=\nu_1+\frac 1{\nu_2+\frac 1{\nu_3+...}},
\ee
where $\nu_i$ are integers. Now, in order to describe strings which satisfy
(\ref{restrn}), one defines series of real numbers $p_i$ and series of integers $m_i$ and $y_i$ as follows:
\bea{notations}
&&p_0=\pi/\lambda,\quad p_1=1,\quad p_i=p_{i-2}-p_{i-1}\nu_{i-1},\nonumber\\
&&m_0=0,\quad m_i=\sum\limits_{k=1}^i\nu_k,\\
&&y_{-1}=0,\quad y_0=1,\quad y_1=\nu_1,\quad y_i=y_{i-2}+\nu_{i}y_{i-1}.\nonumber
\eea
The order and parity of all strings can be expressed by these series:
\bea{ordpar}
&&n_j=y_{i-1}+(j-m_i)y_i\quad\mbox{for}\quad m_i\leq j<m_{i+1},\,\, j=1,2,..,\\
&&v_1=+1,\quad v_{m_1}=-1,\quad v_j=\exp (\pi i[(n_j-1)/p_0]),\nonumber
\eea
where $"[\,\,]"$ is the Gauss' symbol.
Suppose that there are $M_j$ bound states of parity $v_j$ and order $n_j$. Then
we write the coordinate of the center $x^{n_j}_\alpha$ as $x^j_\alpha$. Eqs. (\ref{BAE})
can be written in terms of the centers of strings
\be{sc}
Np_j(x^j_\alpha)= 2\pi J^j_\alpha + \sum\limits _{n=1}^\infty\sum\limits _{\beta=1}^{M_n}\Theta_{jn}(x^j_\alpha-x^n_\beta).
\ee
Here the function $p_j(x)$ is the bare $j-$ string momentum
\be{bpm}
p_j(x)=f(x+\theta/\lambda,n_j,v_j)+ f(x-\theta/\lambda,n_j,v_j),
\ee
and the bare scattering phase between the $j-$ string and the $n-$ string equals
\be{bsp}
\Theta_{jk}(x)=\sum_p
[f(x,p,v_jv_k)+f(x,p+2,v_jv_k)],
\ee
where the sum is over $p=|n_j-n_k|, |n_j-n_k|+2, ..,n_j+n_k-2.$ In eqs. (\ref{bpm}) and (\ref{bsp}) the function
\be{ff1}
f(x,k,v)=i\ln \left[\frac{\sinh\left (\frac{\lambda}{2}(x-i k)-i\frac{\pi}{4}(1-v)\right )}
{\sinh\left (\frac{\lambda}{2}(x+i k)+i\frac{\pi}{4}(1-v)\right )}\right]=2v\tan^{-1}\left[\left (\cot
\left (\frac{\pi n_k}{2p_0}\right )\right)^v\tanh\left (\frac{\pi x}{2p_0}\right )\right ]
\ee
is introduced.

The total momentum of a given configuration is $P=\sum\limits_kp_k(x)$. The energy
can be expressed as $E={\cal A}\sum\limits_kh_k(x),$ where
\be{bareenergy}h_k(x)= \frac d{dx}[f(x-\theta/\lambda,n_k,v_k)-
f(x+\theta/\lambda,n_k,v_k)]\ee
is the bare-particle energy, and the constant 
$$
{\cal A}=\frac{\cosh(2\theta)-\cos(2\lambda)}{\sinh^2\theta}\frac
{\tan\lambda}{\lambda}
$$
is introduced.

The quantities $J^j_\alpha$ are integers or half-integers then $N$ is odd or even respectively. The set $\{J^j_\alpha\}$
defines a unique solution $\{x^j_\alpha\}$  of (\ref{BAE}). A set of numbers $\tilde{J}^j_\alpha$ omitted in $\{J^j_\alpha\}$ defines the holes
in the distribution of $x^j_\alpha$. In the thermodynamic limit, the string "particles" and "holes" are continuously
distributed with densities $\rho_j$ and $\tilde{\rho_j}$. Then eqs. (\ref{BAE}) in the leading order in $1/N$ acquire  the form
of integral equations as follows:
\be{dis1}
b_j(x)=(-1)^{r(j)}(\rho_j(x)+\tilde{\rho_j}(x))+
\sum\limits_k T_{jk}\ast\rho_k(x),
\ee
where
$$
T_{jk}(x)=\frac 1{2\pi}\frac d{dx} \Theta_{jk}(x),
$$
and
$$ b_k(x)=\frac 1{2\pi}\frac
d{dx}[f(x+\theta/\lambda,n_k,v_k)+
f(x-\theta/\lambda,n_k,v_k)].
$$
The convolution "$\ast$" is defined as
$$
f\ast g(x)=\int\limits_{-\infty}^{+\infty}f(x-y)g(y)dy.
$$
The sign-factor $(-1)^{r(j)}=\mbox{sign}(J_\alpha^j-J_{\alpha-1}^j) $ indicates the signature of the derivative
$f^\prime(x,j,v_j)$. It equivalently can be defined from the inequality (see Ref. \cite{takahashi})
$$m_{r(j)}\leq j<m_{r(j)+1}.$$

The energy and momentum are of the form
\bea{energy1}
&&E/N=\sum\limits_{j=1}^{\infty}\int\limits_{-\infty}^{\infty}\left
({\cal A}h_j(x)+2n_j\mu_0H\right)\rho_j(x)dx-\mu_0H,\\
&&P/N=\sum\limits_{j=1}^{\infty}\int\limits_{-\infty}^{\infty}p_j(x)\rho_j(x)dx.
\label{moment1}
\eea
Following to \cite{yang}, we can express the entropy by
\be{entropy1}
S/N=\sum\limits_{j=1}^{\infty}\int\limits_{-\infty}^{\infty}\left
( (\rho_j+\tr_j)\ln(\rho_j+\tr_j)-\rho_j\ln\rho_j-\tr_j\ln\tr_j \right)dx.
\ee

To treat the case $\lambda>\pi/2$, one can use the above mentioned symmetry
$\lambda\rightarrow\pi-\lambda$ and replace $\lambda$ by $\pi-\lambda$ in (\ref{sc})-(\ref{dis1}).
This replacement preserves the previous classification of strings, with the new parameter
$p_0=\frac\pi{\pi-\lambda}$. Eqs. (\ref{dis1}) retain their form, but the bare energy and momentum
change their signs in (\ref{bpm}) and  (\ref{bareenergy}):
\be{chb}
 p_j(x,\lambda)\rightarrow -p_j(x,\pi-\lambda),\qquad h_j(x,\lambda)\rightarrow -h_j(x,\pi-\lambda).
\ee

\subsubsection{The spectral equations}

Let the system is in the state characterized by densities $\rho_j(x)$ and $\tr_j(x)$.
The equilibrium dynamics of the system at some temperature $T$ can be extracted by minimizing the free energy
$F=E-TS$ with respect to independent $\rho_j$. This yields the following non-linear integral equations for
$\eta_j=\tr_j/\rho_j$ (the main spectral equations):
\be{GaudTaka} \ln\eta_j=\frac 1T\left({\cal A}h_j+2n_j\mu_0H\right)+\sum\limits_{k=1}^{\infty}(-1)^{r(k)}T_{j,k}
\ast\ln(1+\eta^{-1}_k).\ee
It is more convenient to deal with a symmetric integral operator
$$A_{jk}(x-y)=(-1)^{r(j)}\delta_{j,k}\delta(x-y)+T_{j,k}(x-y).$$
Its inverse operator has the form
\bea{invA}
&&\hspace{-0.3cm}A_{jl}^{-1}=(-1)^{i+1}\left( \delta_{j,l}\delta(x-y)-s_i[(1-2\delta_{m_{i-1},j})\delta_{j-1,l}+\delta_{j+1,l}]\right),\quad
m_{i-1}\leq j\leq m_i-2,\nonumber\\
&&\,\\
&&\hspace{-0.3cm}A_{jl}^{-1}=(-1)^{i+1}\left( \delta_{j,l}\delta(x-y)-s_i(1-2\delta_{m_{i-1},j})\delta_{j-1,l}-d_i\delta_{j,l}
-s_{i+1}\delta_{j+1,l}\right),\quad j=m_i-1,\nonumber
\eea
where
$$s_i(x)=\frac 1{4p_i}\frac 1{\cosh (\frac {\pi x}{2p_i})},\qquad d_i(x)=
\int\limits_{-\infty}^{\infty}\frac{d\omega}{2\pi}\frac{e^{-i\omega x}}{2}\frac{\cosh[(p_i-p_{i+1})\omega]}{\cosh(p_i\omega)
\cosh (p_{i+1} \omega)}.
$$
This operator acts on $h_j$ and $b_j$ in very simple way:
\bea{propA}
&&\sum_lA_{jl}^{-1}\ast h_l=2\pi\delta _{j,1}\left(s_1(x-\theta/\lambda)-s_1(x+\theta/\lambda)\right)\equiv\varepsilon_0(x).\nonumber\\
&&\,\\
&&\sum_lA_{jl}^{-1}\ast b_l=\delta_{j,1}\left(s_1(x+\theta/\lambda)+s_1(x-\theta/\lambda)\right),\nonumber
\eea
The free energy can be expressed in transparent form
\be{Fenergy} F=-T\int\limits_{-\infty}^{\infty}\left
(s_1(x+\theta/\lambda)+s_1(x-\theta/\lambda) \right)\ln(1+\eta_1)dx.\ee
Here we used (\ref{energy1}), (\ref{GaudTaka}) and the properties (\ref{propA}).

In the zero temperature limit $T\rightarrow 0$ the system goes to its vacuum state. For further analysis of this limit
it is useful to introduce the so-called pseudo-energies by $\eta_j(x)=\exp\left(\frac{\varepsilon_j(x)}T\right)$ and
transform (\ref{GaudTaka}) in terms of these functions as follows:
\be{GaudTaka0}
\varepsilon^+_j(x)={\cal A}h_j(x)+2n_j\mu_0H-\sum\limits_{k=1}^{\infty}(-1)^{r(k)}A_{j,k}
\ast\varepsilon^-_k(x),
\ee
or, using the properties of inverse operator (\ref{propA}),
\bea{vaces}
&&\varepsilon_1(x)={\cal A}\varepsilon_0(x)+s_1\ast\varepsilon_2^+(x),\\
&&\varepsilon_j(x)=s_i\ast[(1-2\delta_{m_{i-1},j})\varepsilon_{j-1}^+(x)+\varepsilon_{j+1}^+(x)],\quad m_{i-1}\leq j\leq
m_i-2\nonumber\\
&&\varepsilon_j(x)=s_i\ast(1-2\delta_{m_{i-1},j})\varepsilon_{j-1}^+(x)+d_i\ast\varepsilon^+_j(x)+
s_{i+1}\ast\varepsilon_{j+1}^+(x),\quad j=m_i-1,\nonumber\\
&&\lim_{j\rightarrow\infty}\frac{\varepsilon_j(x)}{n_j}=2\mu_0H.\nonumber
\eea
Here the dagger means positive parts of corresponding functions:
$$
\varepsilon^+_j(x)=\left \{{\begin{array}{ccc}
                           \varepsilon_j(x), &\mbox{if}&\quad \varepsilon_j(x)\geq 0,\\
                           \,&\,&\,\\
                           0,&\mbox{if}&\quad \varepsilon_j(x)< 0,
                           \end{array} }\qquad \varepsilon^-_j(x)=\varepsilon_j(x)-\varepsilon^+_j(x) \right .
$$

From eqs. (\ref{energy1}) and (\ref{GaudTaka0}) one can represent the energy in a transparent form

$$E=E_0+E_{exc},$$
where
\bea{ven}
E_0/N&&\hspace{-0.5cm}=\sum\limits_j(-1)^{r(j)}\int\varepsilon_j^-(x)b_j(x)dx\\
&&\hspace{0.5cm}=-\int\limits_{-\infty}^{\infty}
\left[s_1(x+\theta/\lambda)+s_1(x-\theta/\lambda)\right]\varepsilon_1^+(x)dx, \label{vacen}\\
E_{exc}/N&&\hspace{-0.5cm}=\sum\limits_j\int\left[\varepsilon_j^+(x)\rho_j(x)-\varepsilon_j^-(x)\tr_j(x)\right]dx.
\label{exen}
\eea
By definition, (\ref{exen}) is non-negative. Therefore, $E_{min}=E_0$ is the energy of vacuum configuration, which can
be realized then the following two conditions are satisfied:
\be{indicates}
\varepsilon_j^+(x)\rho_j(x)=0,\quad \varepsilon_j^-(x)\tr_j(x)=0.
\ee
The expression for vacuum energy (\ref{vacen}) follows from (\ref{dis1}), (\ref{invA}), (\ref{GaudTaka0}) and agrees
with (\ref{Fenergy}) in the limit $T\rightarrow 0$.

\subsection{The case of imaginary $\lambda$ and real $\theta$ $(|J_1|>|J_2|)$}
\subsubsection{The properties of the Bethe Ansatz equations}

In order to ease notations and deal with real spectral parameter and real $\lambda$,
we change $x\rightarrow \frac 1\lambda\varphi$ and $\lambda\rightarrow i\lambda$. The
corresponding Bethe Ansatz equations and all the physical quantities are periodic in each $\varphi_k$ with
a real period $2\pi$.

The string solutions are of the form
\be{strings2}
\varphi^{n,k}_{\alpha}= \varphi^{n}_{\alpha}+i\lambda (n+1-2k),\quad (\mbox{mod}\,\,2\pi),\quad
k=1,2,..,n;
\ee
where $\varphi^{n}_{\alpha}$ is the real part ( the center of string )  and the order $n$ is not bounded from above \cite{gaudin}.

The bare particle momentum $p_n(\varphi)$ and energy $h_n(\varphi)$ are defined like in (\ref{bpm}) and
(\ref{bareenergy}):
\bea{bpm2}
&&p_n(x)=f(\varphi+\theta,n)+ f(\varphi-\theta,n),\\
&&h_k(x)= \frac d{d\varphi}[f(\varphi+\theta,n)-
f(\varphi-\theta,n)].\label{bareenergy2}
\eea
Here the function $f$ is defined as
\be{ff2}
f(\varphi,k)=-i\ln \left[\frac{\sin\frac 12(\varphi-i\lambda k)}
{\sin\frac 12(\varphi+i\lambda k)}\right].
\ee
Due to the mentioned periodicity, one can restrict the interval $\varphi_k\in[-\pi,\pi]$ once fixing the
logarithmic branch in (\ref{ff2}).

The bare two-particle scattering phase $\Theta_{jk}$ is also given by (\ref{bsp}) with the
function (\ref{ff2}).

The densities of the $n$-strings and holes $\rho_n(\varphi)$ and $\tr_n(\varphi)$, defined in the region $[-\pi,\pi]$, obey
the following integral equations:
\be{dis2}
b_j(\varphi)=\rho_j(\varphi)+\tilde{\rho_j}(\varphi)+
\sum\limits_k T_{jk}\ast\rho_k(\varphi),
\ee
where
$$
T_{jk}(\varphi)=\frac 1{2\pi}\frac d{d\varphi} \Theta_{jk}(\varphi),
\qquad b_k(\varphi)=\frac 1{2\pi}\frac
d{d\varphi}[f(\varphi+\theta,n_k,v_k)+
f(\varphi-\theta,n_k,v_k)].
$$
The symbol "$\ast$" means
$$
f\ast g(\varphi)=\int\limits_{-\pi}^{\pi}f(\varphi-\psi)g(\psi)d\psi.
$$

The energy, momentum and entropy of a given configuration $\rho_n(\varphi),\,\tr_n(\varphi)$ are
\bea{energy2}
&&E/N=\sum\limits_{n=1}^{\infty}\int\limits_{-\pi}^{\pi}\left
({\cal A}h_n(\varphi)+2n\mu_0H\right)\rho_n(\varphi)d\varphi-\mu_0H,\\
&&P/N=\sum\limits_{n=1}^{\infty}\int\limits_{-\pi}^{\pi}p_n(\varphi)\rho_n(\varphi)d\varphi.
\label{moment2}\\
&&S/N=\sum\limits_{n=1}^{\infty}\int\limits_{-\pi}^{\pi}\left
( (\rho_n+\tr_n)\ln(\rho_n+\tr_n)-\rho_n\ln\rho_n-\tr_n\ln\tr_n \right)d\varphi.
\label{entropy}
\eea

\subsubsection{The spectral equations}

In this case, the same steps as in previous one are appropriate. First we minimize
the free energy and get the following spectral equations:
\be{GaudTaka1} \ln\eta_n=\frac 1T\left({\cal A}h_n+2n\mu_0H\right)+\sum\limits_{m=1}^{\infty}T_{n,m}
\ast\ln(1+\eta^{-1}_n).
\ee
Then we pass to zero temperature limit, where (\ref{GaudTaka1}) is equivalent to the following
integral equations for pseudo-energies $\eta_n(\varphi)=\exp\left
(\frac{\varepsilon_n(\varphi)}T\right)$:
\be{GaudTaka10}
\varepsilon^+_n(\varphi)={\cal A}h_n(\varphi)+2n\mu_0H-\sum\limits_{m=1}^{\infty}A_{n,m}
\ast\varepsilon^-_m(\varphi),
\ee
where we introduced the operator
$$A_{n,m}(\varphi)=\delta_{n,m}\delta(\varphi)+T_{n,m}(\varphi).$$
Its inverse will have a form
\be{inv1}
A^{-1}_{n,m}(\varphi)=\delta_{n,m}\delta(\varphi)-\frac 1{2\pi}D(\varphi)(\delta_{n,m-1}+\delta_{n,m+1}),
\ee
with
$$D(\varphi)=\sum\limits_{k}\frac{e^{ik\varphi}}{2\cosh(k\lambda)}$$
which is positive valued function with period $2\pi$.
Inverting the system (\ref{GaudTaka10}), one gets:
\bea{vaces1}
&&\varepsilon_1(\varphi)={\cal A}\varepsilon_0(\varphi)+\frac 1{2\pi}D\ast\varepsilon_2^+(\varphi),\\
&&\varepsilon_n(\varphi)=\frac 1{2\pi}D\ast\left (\varepsilon_{n-1}^+(\varphi)+\varepsilon_{n+1}^+(\varphi)\right),\quad
n\geq 2\nonumber\\
&&\lim_{n\rightarrow\infty}\frac{\varepsilon_n(\varphi)}{n}=2\mu_0H.\nonumber
\eea
Here the property
$$A^{-1}_{n,m}\ast h_m(\varphi)=\delta_{1,n}\varepsilon_0(\varphi)=\delta_{1,n}\left (D(\varphi+\theta)-
D(\varphi-\theta)\right)$$
have been used.

This system of coupled equations is simpler than the previous system (\ref{vaces}). In particular, one can immediately
conclude that the functions $\varepsilon_n$ are non-negative for $n\geq 2$. This means that
the vacuum is formed by 1- strings only.

For the energy of given configuration $\rho_n,\,\tr_n$, the equations analogous to (\ref{ven})-
(\ref{exen}) are valid:
$$E=E_0+E_{exc},$$
where
\bea{ven1}
E_0/N&&\hspace{-0.5cm}=\sum\limits_n\int\varepsilon_n^-(\varphi)b_n(\varphi)d\varphi\\
&&\hspace{0.5cm}=-\frac1{2\pi}\int\limits_{-\pi}^{\pi}
\left[D(\varphi+\theta)+D(\varphi-\theta)\right]\varepsilon_1^+(\varphi)d\varphi, \label{vacen1}\\
E_{exc}/N&&\hspace{-0.5cm}=\sum\limits_n\int\left[\varepsilon_n^+(\varphi)\rho_n(\varphi)-\varepsilon_n^-(\varphi)\tr_n(\varphi)\right]d\varphi.
\label{exen1}
\eea

\section{Ground state and excitation spectrum in zero "magnetic" field}

Here we shall study the structure of the ground state and excitation spectrum at $H=0$.

First consider the case of real $\lambda$, (\ref{par1}), in the region $0<\lambda<\pi/2$.
By definition,
$$\eta_j(x)=\tr_j(x)/\rho_j(x)=\exp\left(\frac{\varepsilon_j(x)}T\right).
$$
In the limit $T\rightarrow 0$ survive only the states whose rapidities obey $\varepsilon_j(x)<0$.
Relations (\ref{indicates}) also indicate that, in the ground state, the occupied rapidities obey
$\varepsilon_j<0$. Let us denote this region by $\Delta_j$ (the Dirac sea):
$$\varepsilon_j(x)<0,\quad x\in\Delta_j.$$
According to this, the structure of the ground state is governed by the system of coupled equations
(\ref{vaces}).

It can be shown that in zero "magnetic" field $H=0$ eqs. (\ref{vaces}) have solutions with the following
properties:
\bea{sol}
&&\varepsilon_j(x)=\int\limits_{\varepsilon_1\geq 0}G_j(x-y)\varepsilon_1(y)dy\geq 0,\quad 2\leq j\leq m_1-1,\\
&&\,\nonumber\\
&&G_j(x)=\int\limits_{-\infty}^{\infty}\frac{d\omega}{2\pi}e^{-i\omega x}\frac{\cosh(p_0-j-1)\omega}{\cosh(p_0-2)\omega}\geq 0,\quad 2\leq j\leq m_1-1,\nonumber\\
&&\,\nonumber\\
\label{oddstring}
&&\varepsilon_{m_1}(x)=-s_2\ast\varepsilon_{m_1-1}(x)\leq 0,\\
&&\,\nonumber\\
\label{poch}
&&\varepsilon_j(x)=0,\quad j>m_1.
\eea
These relations constitute the solution of (\ref{vaces}) in terms of unknown function $\varepsilon_1(x)$.
Inserting the integral representation (\ref{sol}) for non-negative $\varepsilon_2$ into
the first equation of (\ref{vaces}),  we arrive to the following integral equation with respect to $\varepsilon_1(x)$:
\bea{mieq}
&&\varepsilon_1(x)={\cal A}\varepsilon_0(x)+\int\limits_{\varepsilon_1\geq 0}R(x-y)\varepsilon_1(y)dy,\\
&&R(x)=s_1\ast G_2(x)=\int\limits_{-\infty}^{\infty}\frac{d\omega}{2\pi}e^{-i\omega x}\frac{\cosh[(p_0-3)\omega]}
{2\cosh[(p_0-2)\omega]\cosh [\omega]}\geq 0.\nonumber
\eea

Here one of course needs to solve eq. (\ref{mieq}). One could assume that the Wiener-Hopf method
would be effective. This method would nicely work in the case when the region $\varepsilon_1\geq 0$
is spanned from infinity to some finite point. Our analysis shows that this is not the case. 
Then We employed both perturbative Wiener-Hopf and numerical methods. 
In order to be shorter, we will present the detailed analysis and related results elsewhere.
Meanwhile we bring here some speculative analysis which make the general picture rather clear.

As defined by (\ref{propA}), the so-called disturbance function $\varepsilon_0$ is an odd function 
with positive values at
positive arguments, $\varepsilon_0(x)>0,\,\,x>0$. This means that $\varepsilon_1$ is also positive
at non-negative values of $x$: $\varepsilon_1(x)>0,\,\, x>0\,\,({\cal A}>0)$. Further analysis of eq. (\ref{mieq})
shows that $\varepsilon_1(x)$ takes negative values in some region in $(-\infty,0)$. This region which we denoted by $\Delta_1$,
is the Dirac sea of $1$- strings. Furthermore, $\varepsilon_{m_1}$ is non-positive within the whole
real axis $-\infty<x<+\infty$. So the Dirac sea of $m_1$- strings $\Delta_{m_1}$ is $(-\infty,+\infty)$.
The rest pseudo-energies are non negative. Thus the vacuum is filled by the $1$ and $m_1$ strings:
\bea{vdist}
\rho^0_1(x)\geq 0,\quad\tilde{\rho}^0_1(x)\geq 0,
&&\Delta_1\subset(-\infty,0),\nonumber\\
\rho^0_{m_1}(x)\geq 0,\quad\tilde{\rho}^0_{m_1}(x)=0,
&&\Delta_{m_1}=(-\infty,+\infty),\\
\rho^0_j(x)= 0,\quad j\neq 1,m_1,
&&\Delta_j=\emptyset,\nonumber \\
&&\,\nonumber\\
 \Delta=&&\hspace{-0.5cm}\Delta_1\oplus \Delta_{m_1}.\nonumber
\eea
where $\rho^0,\,\,\tilde\rho^0$ are vacuum distribution functions, $\Delta$ is the whole Dirac sea.
Integrating (\ref{dis1}) for $j=m_1$ we get that the vacuum configuration is $M^0_1+M^0_{m_1}=N,$
$M^0_j=0,\,\,j\neq 1,m_1$, where $M^0_i$ are numbers of vacuum $i$-strings. So we have the half-filled band and $S_z=0$,
i.e., we deal with an antiferromagnetic case.

Let us turn to considerations of excitation  spectrum. The possible excited states can be formed
by adding $l_k$ excited $k$-strings with rapidities $z^k_\alpha$,
$\alpha =1,..,l_k$ and subtracting $v_j$ vacuum $j$- strings with the rapidities $z_\alpha^{h,k}$.
This state will have a total spin
\be{exspin}
S_z= \sum\limits_j n_jv_j -\sum\limits_kn_kl_k\geq 0,
\ee
and energy
\be{phen}
{\cal E}=\sum\limits_k\sum\limits_{\alpha=1}^{l_k}\varepsilon_k(z^k_\alpha)-
\sum\limits_j\sum\limits_{\alpha=1}^{v_j}\varepsilon_j(z_\alpha^{h,j}),\qquad
z^k_\alpha\in (-\infty;+\infty)\setminus\Delta_k,\quad z_\alpha^{h,j}\in\Delta_j.
\ee

The BAE poses restrictions on the possible combinations of excitation parameters $\{z_\alpha^{k}\}$ and $\{z_\alpha^{h,j}\}$.
We miss here details of the treatment of this problem, which would involve the analysis of Eqs. (\ref{sc}) in the next to
leading order (\ref{dis1}), i.e., the $1/N$ corrections to (\ref{dis1}). This is based on the known methods \cite{devegawoy,kar,woy}
and will be reported elsewhere. Instead we bring here some results for lowest-lying excitations with 
$|M_i-M^0_i|\leq 1$.
\begin{itemize}
\item $M_i=M^0_i;$ The particle-hole excitations: 
\be{ll1}E=-\varepsilon_1(z^{h,1})+\varepsilon_1(z^1),\quad S_z=0.\ee
\item $M_1=M^0_1-1,\,\,M_i=M^0_i,\,\, i>1;$ The two-hole excitations: 
\be{ll2}E=-\varepsilon_1(z^{h,1}_1)-\varepsilon_1(z^{h,1}_2),\quad S_z=1.\ee
\item $M_{m_1}=M^0_{m_1}-1,\,\,M_i=M^0_i,\,\, i\neq m_1;$ The particle-hole excitations: 
\be{ll3}E=-\varepsilon_1(z^{h,1})+\varepsilon_1(z^{1}),\quad S_z=1.\ee
\item $M_1=M^0_1-1,\,\,M_{m_1}=M^0_{m_1}+1,\,\,M_i=M^0_i,\,\, i\neq 1,m_1;$ The two-hole excitations: 
\be{ll4}E=-\varepsilon_1(z^{h,1}_1)-\varepsilon_1(z^{h,1}_2),\quad S_z=0.\ee
\item $M_1=M^0_1+1,\,\,M_{m_1}=M^0_{m_1}-1,\,\,M_i=M^0_i,\,\, i\neq 1,m_1;$ The two-particle excitations: 
\be{ll5}E=\varepsilon_1(z^{1}_1)+\varepsilon_1(z^{1}_2),\quad S_z=0.\ee
\end{itemize}
All the rest excitations have $|M_i-M^0_i|>1$ for some $i$. Then, it is possible to have some combinations of
excitations given above, as well as a new ones. For example, when $p_0>M^0_{m_1}-M_{m_1}>\frac{p_0}2\,(>1)
,\,\,M_i=M^0_i,\,\, i\neq m_1,$ together with other possible excitations  it appears a new branch of two $m_1$- hole 
excitation with the energy 
\be{ll6}E=-\varepsilon_{m_1}(z^{h,2}_1)-\varepsilon_{m_1}(z^{h,2}_2).\ee

Notice that in any case the particle-hole excitation (\ref{ll1}) can be repeatedly excited: they can appear in any number as well as 
in combination with any other excitations.

Besides these branches of the spectrum, there also exist bound states of magnons whose energies
are given by $\varepsilon_j,\,j=1,2,..,m_1-1,$ (\ref{sol}). These bound states are single-parametric,
spin-singlet elementary excitations. Indeed, to construct a single bound state of $s$ magnons,
we must choose
$$ l_k=\delta_{k,s},\quad \rho_1(x)=0,\,\,x\not\in\Delta_1,\quad \tr_1(x)=0,\,\,x\in\Delta_1,\quad \tr_{m_1}=0.$$
Then the $j=1$ and $j=m_1$ eqs. (\ref{dis1}) acquire the form
\bea{exdis}
&&b_1(x)=\rho_1(x)+\tilde{\rho_1}(x)+
T_{11}\ast\rho_1(x)+T_{1m_1}\ast\rho_{m_1}(x)+\frac 1N T_{1s}(x-z_s)\\
&&b_{m_1}(x)=-\rho_{m_1}(x)+T_{m_11}\ast\rho_1(x)+T_{m_1m_1}\ast\rho_{m_1}(x)
+\frac 1N T_{m_1s}(x-z_s),\nonumber
\eea
where $z_s$ parameterizes the excitation. From the last equation it follows that this configuration
has a zero spin,
$$S_z=v_1+v_{m_1}-s l_s=0.$$
It can be shown that (\ref{exdis}) has a unique solution for $z_s\in(-\infty;+\infty)$. It constitutes
elementary excitation with the energy
\be{brex}
  {\cal E}=\varepsilon_{s}(z_s).
\ee
As seen from (\ref{exen}) and(\ref{poch}), strings with $j>m_1$ have vanishing energies and become
important only when one classifies excited states with respect to the total spin $S_z$.

Due to the properties of spectral functions $\varepsilon$ (\ref{sol})-(\ref{poch}), 
excitation energies can take any positive values close to zero, then excitation
parameters $\{z^1\}$ and $\{z^h\}$ come close to Fermi points or $z_s\rightarrow\pm\infty$, i. e., 
the spectrum is massless, as it was expected.

In the region $\pi/2<\lambda<\pi,$ we use transformation (\ref{chb}) which produces a mapping
of this region onto the previous one.  From (\ref{chb}) and due to the fact that $s_i(x)$ and $d_i(x)$ in
(\ref{invA}) are even functions of $x$ whereas $h_j(x)$ and
$\varepsilon_0(x)$ in (\ref{propA}) are odd functions of $x$, one gets that all of our solutions and results remain valid
in this region also, when we change $\lambda$ by $\pi-\lambda$ and the variable $x$ by $-x$,
in the corresponding functions.

In the second case (\ref{par2}) of imaginary $\lambda$ we deal with the spectral system
(\ref{vaces1}). It can be easily shown that
\be{sol2}
\varepsilon_n=\int\limits_{-\pi}^{\pi}\frac{\sinh[(n-1)\lambda]}{\cosh[(n-1)\lambda]
-\cos(\varphi-\psi)}\varepsilon_1^+(\psi)d\psi>0\quad\mbox{for}\quad n\geq 2
\ee
obeys (\ref{vaces1}) if $\varepsilon_1(\varphi)$ is solution to the integral equation
\be{inteq2}
\varepsilon_1(\varphi)={\cal A}\varepsilon_0(\varphi)+\int\limits_{-\pi}^{\pi}
r(\varphi-\psi)\varepsilon_1^+(\psi)d\psi,
\ee
wher the real positive valued kernel $r(\varphi)=\frac 1{2\pi}\sum\limits_k\frac{e^{-|\lambda k|}}
{2\cosh(\lambda k)}e^{ik\varphi}$is periodic with $2\pi$. As in the previous case, from this integral
equation we conclude that $\varepsilon_1(\varphi)$ takes negative values in some region $\Delta\subset(-\infty,0)$.
The vacuum is formed by 1-strings filled with rapidities from $\Delta_1$; the vacuum densities 
satisfy
$$ \rho_1^0(\varphi)=0,\,\,\varphi\not\in\Delta_1,\quad\tr_1^0=0,\,\,\varphi\in\Delta_1,$$
$$\rho_n^0(\varphi)=0,\,\,\varphi\in [-\pi,\pi],\,\, n=2,3,... .$$
By the integration of (\ref{dis2}) for $n=1$ one gets for the vacuum configuration 
\be{vaccon}
M_1+\frac 12\tilde{M}_1=N,\qquad M_1,\tilde{M}_1\neq 0,\qquad M_n=0,\quad n\geq 2.
\ee 
This means that 
our vacuum in this case has spontaneous spin $S_z\neq 0$. The band-filling is $\frac {M_1}{2N}<1/2$
(the exact value of band filling can be extracted from the exact solution of (\ref{inteq2}) and
 (\ref{dis2}) only ).

The excitation spectrum includes massless 1-string particle-hole excitations with energies
\be{exen2}{\cal E}=\sum\limits_p\varepsilon_1(\psi_p)-
\sum\limits_q\varepsilon_1(\psi_q^h),\qquad
\psi_p\not\in\Delta_1,\quad \psi_q^{h}\in\Delta,
\ee
and massive breathers of order $n,\,\,n\geq 2$ with energies
\be{exbren2}{\cal E}_n=\sum\limits_p\varepsilon_n(\psi_p).
\ee

\section{Some Special Limits}
\subsection{$J_1=0,\quad J_2\neq 0$}

This case is also in the region (\ref{par1}). It corresponds to the limit
 $\theta\rightarrow \pm i\infty$, when $ J_1=0$ and $J_2=\mp i\tan\lambda.$
Consider the integral equation (\ref{mieq}) for $\theta\gg 1$ (we are changing
$\theta\rightarrow i\theta$). By the definition (\ref{propA}), $\varepsilon_0$ depends on
the combinations $x\pm\theta/\lambda$ and is essentially non-zero in the vicinity of $\pm\theta/\lambda.$
One can expect the solution $\varepsilon_1$ has the same property and expand (\ref{mieq})
near to these points: the mutual influence of these regions will be of order $O(e^{-\pi|\frac\theta\lambda|})$.
Then (\ref{mieq}) goes to the following two decoupled integral equations:
\bea{2ind}
&&\varepsilon_1^R(x)=\frac{\pi\tan\lambda/\lambda}{\cosh(\frac\pi 2x)}+\int\limits_{\varepsilon_1^R\geq 0}R(x-y)\varepsilon_1^R(y)dy,\\
&&\varepsilon_1^L(x)=-\frac{\pi\tan\lambda/\lambda}{\cosh(\frac\pi 2x)}+\int\limits_{\varepsilon_1^L\geq 0}R(x-y)\varepsilon_1^L(y)dy,\nonumber
\eea
where $\varepsilon_1^L(x)=\varepsilon_1(x-\theta/\lambda)$, $\varepsilon_1^R(x)=\varepsilon_1(x+\theta/\lambda)$.

By their forms, Eqs. (\ref{2ind}) are the same as corresponding integral equation for the 
spectral function $\varepsilon_1$ of Heisenberg XXZ model (see for example \cite{takahashi}): from 
this naive point of view, here we have two samples of XXZ chains with different signs of 
coupling constants.

It is natural to try to confirm this double chain structure starting from the Hamiltonian, 
which on the other hand would be seen a self-consistency check for our results. 
In the case under consideration, the Hamiltonian (\ref{spinham}) takes the form
\bea{ham}
H&&\hspace{-0.5cm}=\sum\limits_{j=1}^N\sum\limits_{s=0,1}\left \{{(-1)^s}/2 \left[  \sigma_{j,s}^x \sigma_{j+1,s}^x+  \sigma_{j,s}^y
\sigma_{j+1,s}^y-  \sigma_{j,s}^z \sigma_{j+1,s}^z\right]\right .\\
&&\,\nonumber\\
&&\hspace{3cm}+\tan\lambda \, i\left .\left[ \sigma_{j,s+1}^z( \sigma_{j,s}^+ \sigma_{j+1,s}^- -  \sigma_{j,s}^- \sigma_{j+1,s}^+)\right]
\right \}\nonumber
\eea
Being rewritten as follows:
\be{ef}
H=\frac1{\cos\lambda}\sum\limits_{j=1}^N\sum\limits_{s=0,1}(-1)^s \left[\left (\exp\{(-1)^si\lambda\sigma_{j,1-s}^z\}
\cdot\sigma_{j,s}^+\sigma_{j+1,s}^- +h.c.\right)-\frac{\cos\lambda}2 \sigma_{j,s}^z \sigma_{j+1,s}^z\right],
\ee
it reveals the underlying double chain structure, which is more evident under the following gauge-like unitary transformation:
\bea{bham}
H&&\hspace{-0.5cm}=\frac{1}{\cos\lambda}e^{-i\frac\lambda 2\sum\limits_{ N\geq n>k\geq 1}\sigma_{n,0}^z\sigma_{k,1}^z}
\left [H^{xxz}_0+ \left(e^{i\lambda\sum\limits_{k=1}^N\sigma_{k,1}^z} \sigma_{N,0}^+ \sigma_{1,0}^- + h.c.\right)
\right .\nonumber\\
&&\,\nonumber\\
&&\hspace{4cm} \left . -H^{xxz}_1- \left(e^{-i\lambda\sum\limits_{n=1}^N\sigma_{n,0}^z} \sigma_{N,1}^+ \sigma_{1,1}^-
+ h.c.\right)\right ]e^{i\frac\lambda 2\sum\limits_{N\geq n>k\geq 1}\sigma_{n,0}^z\sigma_{k,1}^z};
\nonumber
\eea
where
$$
 H^{xxz}_s=H^{xxz}_{s,\, open}=\sum\limits_{l=1}^{N-1} \sigma_{l,s}^+\sigma_{l+1,s}^- +\sigma_{l,s}^-\sigma_{l+1,s}^+
 +\frac12\cos\lambda\,\sigma_{l,s}^z\sigma_{l+1,s}^z.
$$
So the system can be described by the unitarily equivalent Hamiltonian
\be{btham}
\tilde{H}=\frac1{\cos\lambda} \left \{ H^{xxz}_0+ (e^{i\lambda\sum\limits_{k=1}^N\sigma_{k,1}^z} \sigma_{N,0}^+ \sigma_{1,0}^- + h.c.)
 -H^{xxz}_1- (e^{-i\lambda\sum\limits_{n=1}^N\sigma_{n,0}^z} \sigma_{N,1}^+ \sigma_{1,1}^-
+ h.c.)\right\}
\ee 
By the further application of Bethe Ansatz technique, one can diagonalize this Hamiltonian. 
Then one parameterizes the eigenvalues of (\ref{ham}) by means of two sets of complex 
rapidities $\{u_{s}^i\}$, as attached to the chains $H^{xxz}_s,$ $s=0,1$. The eigenvalue problem
leads to the Bethe Ansatz equations  
\bea{couplBAE}
&&\left[\frac{\sinh\left(\frac\lambda 2(u_1^k-i)\right)}{\sinh\left(\frac\lambda 2(u_1^k+i)\right)}\right]^N=
-e^{i\lambda(N-2M_0)}\prod\limits_{i=1}^{M_1}
\frac{\sinh\left(\frac\lambda 2(u_1^k-u^i_1-2i)\right)}{\sinh\left(\frac\lambda 2(u^i_1-u^k_1-2i)\right)}
\nonumber\\
&&\left[\frac{\sinh\left(\frac\lambda 2(u^k_0-i)\right)}{\sinh\left(\frac\lambda 2(u^k_0+i)\right)}\right]^N=
-e^{-i\lambda(N-2M_1)}\prod\limits_{i=1}^{M_0}
\frac{\sinh\left(\frac\lambda 2(u^k_0-u^i_0-2i)\right)}{\sinh\left(\frac\lambda 2(u^i_0-u^k_0-2i)\right)}.
\eea
The phase factors reflect the twisted boundary  terms in the Hamiltonian $\tilde{H}$. On the
other hand, these phase factors reflect an interaction between the two sets; one could assume, that the $s=0$ spins are inside a "gauge field"
generated by the $s=1$ spins and vice versa.
The energy eigenvalues are given by the formula
\be{couplenergy}
E=E^0+E^1,\quad E^{s}(\{u^i_s\})=(-1)^s2\tan\lambda\sum\limits_{i=1}^{M_{s}}
\frac{\sin\lambda}{\sinh(\frac\lambda 2u^i_{s}-i\frac\lambda 2)\sinh(\frac\lambda 2u^i_{s}+i\frac\lambda 2)},
\ee
which does not affect the mentioned interaction explicitly.
 
 Let us mention that the same equations one would get from BAE (\ref{BAE}). Indeed, in the limit
$\theta\gg 1$, ($\theta\rightarrow i\theta$), the roots of BAE (\ref{BAE}) can be divided into two groups gathered 
close to the points $\pm \theta/\lambda$: $u^{0}=u+\theta/\lambda$ and $u^{1}=u-\theta/\lambda$. Taking the limit
$\theta\rightarrow\infty$ in the BAE (\ref{BAE}), one would get (\ref{couplBAE}).

It is intuitively clear that, in the thermodynamic limit $N\rightarrow\infty$, boundary 
terms does not affect the bulk behavior and the Hamlitonian (\ref{ham}) will have the same 
critical properties as two non-interacting $XXZ$- chains with usual periodic boundary conditions
and opposite signs of coupling constants. This is exactly what we had in (\ref{2ind}); the $s=0$ and $s=1$ chains 
correspond to $L$ and $R$ sectors respectively. 

The interaction effect is of the next to leading order (actually of order $1/N^2$). 
It can be studied using the known finite size correction methods \cite{devegawoy,kar,woy}.
The idea is to consider two $XXZ$ chains, each with additional phase shifts in the Bethe 
Eqs. (\ref{sc}). The phase shift of one of the chains is described by the total spin of another 
chain. This requires rather long calculations. We bring here only the result for
the energy difference between the ground state and the one with
$n^R_s$ ($n^L_s$) particles added at the right (left) Fermi points of the $s=0,1$ chains:
\be{EH}
\Delta E=\frac{2\pi}{4N^2}\sum_sv_{0,s}\left[(\frac1{\xi_s^2}+\xi^2_{1-s}\delta^2_{1-s})\cdot N_s^2+\xi_s^2\cdot J_s^2+
2(-1)^s\delta_s\xi_s^2\cdot J_sN_{1-s}+ K^R_s+K^L_s-\frac13\right],
\ee
where $v_0$ is the Fermi velocity and $\xi$ is the dressed charge introduced in \cite{kor}, 
\be{fvdc}
v_{0,s}=\frac\pi{2\lambda_s}\sin\lambda_s,\quad\xi^2_s=\frac\pi{2(\pi-\lambda_s)};\quad
\lambda_0\equiv\lambda,\,\,\lambda_1\equiv\pi-\lambda.
\ee
and $K^{L,R}$ are some integer quantum numbers.
The interaction parameter $\delta$ comes from the shifted phases. It is defined as
\be{intph}
N_{1-s}\delta_s=\left\{\frac\lambda{2\pi}N_{1-s}\right\},\quad s=0,1,
\ee
where $\{x\}$ is the fractional part of $x$. Quantum numbers $K$ are
unimportant here, because are factored from the remaining interaction.
We also used notations $N_s=n^R_s+n^L_s$, $J_s=n^R_s-n^L_s$. Thus we get the 
"effective Luttinger liquid" description of our motel \cite{haldane}.

A similar integrable model containing two species of $XXZ$ chains with the same signs of 
coupling constants is discussed by Schulz and Shastry in \cite{shsh}. There the equations equivalent 
to (\ref{couplBAE}) are derived in terms of momenta of Bethe excitations. The fact that the
hopping parameters are of the same signs makes the physics different from our case.

Following Ref. \cite{gnt}, we finish this section with the bosonization study of (\ref{ham}).
In order to obtain a right continuum theory, we first apply the Jordan-Wigner transformation 
along the zigzag links. The continuum fermionic Hamiltonian is obtained by the replacement 
$\psi_{n,s}\rightarrow\sqrt {a_0} \,(R_s(x)+(-1)^nL_s(x))$,
$x=na_0$, where $\psi_{n,s}$ is the discrete Fermion field, $R(x)$ and $L(x)$ are the continuous  right and left Dirac 
fields respectively, $a_0$ is the lattice spacing. In the leading order we get:
\bea{FH}
&&H=H_0+H_{int},\nonumber\\
&&\,\nonumber\\
&&H_0=v_0i\,\int dx\sum_s\left(R^+_s(x)\partial R_s(x)-L^+_s(x)\partial L_s(x)\right),\nonumber\\ 
&&\,\nonumber\\
&&H_{int}=v_0g\,\int dx\sum_s(-1)^s\left\{\frac1{2a_0}(J^s_R-J^s_L)+(J^s_R+J^s_L)(J^{1-s}_R-
J^{1-s}_L)+(R^+_sL_s+L^+_sR_s)^2\right\}.\nonumber 
\eea
Here $J^s_R=R^+_sR_s$, $J^s_L=L^+_sL_s$ are the right and left fermion currents respectively, $v_0$
is the Fermi velocity and $g=2\cot\lambda\tanh\theta$ is the coupling constant. Then, using the Bosonization prescriptions, we arrive at the 
bosonic Hamiltonian
\be{BH}
H=v_0\int dx\left\{(1-\frac g{2\pi})(\Pi_0^2+(\partial\Phi_1)^2)+(1+\frac g{2\pi})(\Pi_1^2+(\partial\Phi_0)^2)
+\frac g\pi(\Pi_0\partial\Phi_1-\Pi_1\partial\Phi_0)\right\}.
\ee
Here we introduced Bose fields $\Phi_s(x)$ and canonical conjugate momenta $\Pi_s(x)$ with the 
commutations    
$$[\Phi_s(x),\Pi_{s^\prime}(y)]=i\delta_{ss^\prime}\delta(x-y),$$
In deriving (\ref{BH}) we have dropped strongly irrelevant interaction terms like 
$\cos(4\sqrt\pi\,\Phi_s(x))$. It is not hard to check that in the first order in small 
$\lambda$ the Hamiltonian (\ref{BH}) gives the energy difference equivalent to (\ref{EH}).
The latter allows also to extract an exact values for Fermi velocities.
  
The canonical transformation
\bea{ct}
\partial Q_+=\frac 1{\sqrt K_+}\frac{\partial\Phi_0+\Pi_1}{\sqrt 2},&& 
\partial Q_-=\frac 1{\sqrt K_-}\frac{\partial\Phi_1+\Pi_0}{\sqrt 2},\nonumber\\
&&\,\nonumber\\
P_+={\sqrt K_+}\frac{\Pi_0-\partial\Phi_1}{\sqrt 2},&& 
P_-={\sqrt K_-}\frac{\Pi_1-\partial\Phi_0}{\sqrt 2},\nonumber\\
&&\,\nonumber\\
&K_{\pm}=\sqrt{1\pm g/\pi},\nonumber
\eea
brings (\ref{BH}) into the Gaussian form
\be{Gf}
H=v_0K_+\int dx\left(P_+^2+\partial Q_+^2\right) + v_0K_-\int dx\left(P_-^2+\partial Q_-^2\right).
\ee
We thus see that the interaction between the two chains effects on the renormalization of 
Fermi velocities only.

\subsection{$J_1=J_2$}

This special limit is a crossing of two regions (\ref{par1}) and (\ref{par2}). 
It takes place when $\lambda\rightarrow 0$ and $\theta\rightarrow 0$ with
finite ratio $\frac\theta\lambda=const$. In order to consider this limit, it is convenient to redefine
$\theta\rightarrow i\lambda\theta$ in the region (\ref{par1}), and
$\lambda\rightarrow i\lambda$, $\theta\rightarrow -i\lambda\theta$ and $\varphi\rightarrow \lambda \varphi$,
in the region (\ref{par2}) (here $\varphi$ is the rapidity variable introduced in the section (2.2)). In both cases, we will have
$$J_1=J_2=-i\frac 1{\theta}.$$

Consider this limit within (\ref{par1}). Here one has  $p_0\rightarrow\infty$ and
$\nu_1=m_1\rightarrow\infty$. This makes the system of coupled spectral equations (\ref{vaces}) 
simpler; the pseudo-energies $\varepsilon_j$ become non-negative for $j\geq 2$, like in the 
case of (\ref{vaces1}). Then (\ref{GaudTaka0}) for $j=1$ goes into the integral equation for $\varepsilon_1(x)$ as follows:
\be{mieq1}
\varepsilon_1(x)={\cal
A}_0h(x)+\int\limits_{\varepsilon_1\leq
0}T_0(x-y)\varepsilon_1(y)dy,
\ee
where ${\cal A}_0=2\left(1+1/{\theta^2}\right),\,\,T_0(x)=\frac{1}{2\pi}\frac{1}{(x/2)^2+1},\,\,h(x)=
\frac{2}{(x-\theta)^2+1}-\frac{2}{(x+\theta)^2+1}$. The rest
$\varepsilon_j$ for $j\geq 2$ can
be expressed in terms of the solution of (\ref{mieq1}) through
(\ref{sol}), where
$$G_j(x)=\int\limits_{-\infty}^{\infty}\frac{d\omega}{2\pi}e^{-i\omega
x}e^{-(j-1)|\omega|}=\frac{1}{2\pi}\frac{2(j-1)}{x^2+(j-1)^2}.$$

Let us mention that one in principle
could take the limit in (\ref{mieq}). Then of course the obtained integral equation will be equivalent
to (\ref{mieq1}); instead of our case, it will contain an integral over the region $\varepsilon_1\geq 0$.

The same equation could be found in the region (\ref{par2}). Indeed, the rescaling 
$\varphi\rightarrow \lambda \varphi$ destroys the periodicity  $\varphi\rightarrow  \varphi+2\pi$ and the new
$\varphi$ varies from $-\infty$ to $+\infty$. Taking this in account, one can 
make sure that (\ref{GaudTaka10}) for $n=1$ goes into (\ref{mieq1}). This proves that we indeed
get the same pictures coming from two different regions (\ref{par1}) and (\ref{par2}).

From (\ref{mieq1}) one can conclude that $\varepsilon_1$ takes negative values in some region
$\Delta_0\subset (-\infty,0)$. Thus the vacuum in this case is filled by 1-strings with the
rapidities from this region; the corresponding vacuum densities with the properties
\be{fxprop}
\rho_v(x)=0,\quad x \not \in \Delta_0,\qquad \tr_0(x)=0,\quad x \in \Delta_0
\ee
obey the integral equation 
\be{dis11}
b(x)=\rho_v(x)+\tilde{\rho_v}(x)+ \int\limits_{\Delta_0}
T_{0}(x-y)\rho_v(y)dy, 
\ee 
where $b(x)=\frac{1}{2\pi}\left[\frac{2}{(x+\theta)^2+1}
+\frac{2}{(x-\theta)^2+1}\right ]$. This integral equation together with (\ref{fxprop})
uniquely defines the vacuum densities. The vacuum configuration have properties very similar
to the one from the section (2.2), given in (\ref{vaccon}).

Another phase transition occurs when one in addition sends $\theta\rightarrow\pm\infty$.
It corresponds to $J_1=J_2=0$. This case can be treated similar to the special limit 
from the previous section. It can be checked that (\ref{mieq1}) splits into two integral equations
which are equivalent to corresponding spectral equations of isotropic $XXX$-chain. The two integral equations
will differ by opposite signs of the free term (disturbance function), which means that the corresponding isotropic
chains are one ferromagnetic and another antiferromagnetic. The same follows from the Hamiltonian
straightforwardly, when one puts $J_1=J_2=0$. This is another consistency proof of our results.  

\section{Summary and Conclusion}

To summarize we have applied the method introduced by Gaudin in \cite{gaudin} and Takahashi and 
Suzuki in \cite{takahashi} to the exactly solvable two-leg ladder model with zigzag like interaction 
constructed in \cite{ararub}. We have established two distinct zero temperature phases, corresponding to 
two regions in coupling constant space $|J_1/J_2|<1,$ (\ref{par1}) and $|J_1/J_2|> 1,$ (\ref{par2}). We considered the model without
magnetic field. When $|J_1/J_2|< 1,$ we found the model has gapless excitations above the 
antiferromagnetic ground state formed by filling all the possible 1-strings (even magnons) 
with rapidities from some region $\Delta_1\subset (-\infty,0)$ and $m_1$-strings (odd magnons)
with rapidities from $(-\infty,+\infty)$. So this phase is conformal. In the region (\ref{par2}),
we have seen the model possesses a ferromagnetic ground state with gapless 
particle-hole excitations and massive breathers of length $n\geq 2.$ The intersection of these
two phases $|J_1|=|J_2|$ was considered in the last section. In the last section, we have considered also
the special limit when $J_1=0$. We have shown that in this case our spectral equations are 
equivalent to analogous equations for two $XXZ$ chains with opposite signs of coupling constants. 
This double chain structure was confirmed by a unitary
transformation of the Hamiltonian. We would like to mention that 

It naturally arises a question about the scaling dimensions of fields when the conformal phase 
occurs. These calculations could be done by use of the method introduced in \cite{woy2} and
\cite{k+f}.  

The problem which did not allow to get a precise values of Fermi points, Fermi velocities,
momenta of states, as well as conformal dimensions is to solve a Fredholm type equations
as (\ref{mieq}) and (\ref{inteq2}) are. As we already mentioned, the perturbative Wiener-Hopf method
could of course be effective, though it does not usually give an explicit solutions. The
numerical analysis is also appropriate. We plan to present this analysis and related results 
elsewhere.

\vspace{0.5cm}

\section*{Acknowledgement}

\noindent

A.S. is thankful D.Arnaudon and P.Sorba for many interesting
discussions. V.M. would like to express his gratitude to A.Nersesyan and R.Flume 
for valuable discussions and hospitality at ICTP and Bonn university.
The work of A.S. was supported in part by SNF SCOPE grant and INTAS
grant 00390. The work of V.M. was supported by SNF SCOPE grant,
Volkswagen Foundation  and INTAS grant 00561.

\end{document}